\documentclass[iop]{emulateapj}

\journalinfo{To Appear in ApJ}
\submitted{Received 2011 February 17; accepted 2011 June 23}

\shorttitle{{\it KEPLER} EC HOSTS ARE METAL RICH}
\shortauthors{SCHLAUFMAN \& LAUGHLIN}

\begin{document}

\title{{\it KEPLER} EXOPLANET CANDIDATE HOST STARS ARE PREFERENTIALLY
METAL RICH}

\author{Kevin C. Schlaufman\altaffilmark{1} and Gregory Laughlin}
\affil{Astronomy and Astrophysics Department, University of California,
       Santa Cruz, CA 95064}
\email{kcs@ucolick.org and laughlin@ucolick.org}

\altaffiltext{1}{NSF Graduate Research Fellow}

\begin{abstract}

\noindent
We find that {\it Kepler} exoplanet candidate (EC) host stars are
preferentially metal-rich, including the low-mass stellar hosts of
small-radius ECs.  The last observation confirms a tentative hint that
there is a correlation between the metallicity of low-mass stars and
the presence of low-mass and small-radius exoplanets.  In particular,
we compare the $J-H$---$g-r$ color-color distribution of {\it Kepler}
EC host stars with a control sample of dwarf stars selected from the
$\sim\!\!150,000$ stars observed during Q1 and Q2 of the {\it Kepler}
mission but with no detected planets.  We find that at $J-H = 0.30$
characteristic of solar-type stars, the average $g-r$ color of stars
that host giant ECs is 4-$\sigma$ redder than the average color of the
stars in the control sample.  At the same time, the average $g-r$ color
of solar-type stars that host small-radius ECs is indistinguishable from
the average color of the stars in the control sample.  In addition, we
find that at $J-H = 0.62$ indicative of late K dwarfs, the average $g-r$
color of stars that host small-radius ECs is 4-$\sigma$ redder than the
average color of the stars in the control sample.  These offsets are
unlikely to be caused by differential reddening, age differences between
the two populations, or the presence of giant stars in the control sample.
Stellar models suggest that the first color offset is due to a 0.2
dex enhancement in [Fe/H] of the giant EC host population at $M_{\ast}
\approx 1~M_{\odot}$, while Sloan photometry of M~67 and NGC~6791 suggests
that the second color offset is due to a similar [Fe/H] enhancement of
the small-radius EC host population at $M_{\ast} \approx 0.7~M_{\odot}$.
These correlations are a natural consequence of the core-accretion model
of planet formation.

\end{abstract}


\section{Introduction}

The groundbreaking early results of NASA's {\it Kepler} mission
\citep{bor10a,koc10a,jen10a,cal10,bry10,bat10a,bat10b,haa10,jen10b}
have spectacularly added to our understanding of
the frequency and properties of planets in the Galaxy
\citep{bor11a,bor11b,for11,lis11b,moo11,how11,lat11}.  Eight
planetary systems discovered by {\it Kepler} have already been
confirmed by radial velocity or transit timing: four hot Jupiters
\citep{koc10b,dun10,lat10,jen10c}; two Saturn-mass planets and
a super-Earth in the Kepler-9 system \citep{holma10}; a Neptune-mass
planet Kepler-4b \citep{bor10b}; Kepler-10b, the smallest exoplanet yet
known \citep{bat11}; and the closely-aligned six transiting planet system
Kepler-11 \citep{lis11a}.

The properties of the exoplanet candidate (EC) population discovered
by {\it Kepler} will provide strong constraints on models of
planet formation, especially for low-mass stars and small planets.
It is well established that metal-rich solar-type stars are more
likely to host giant planets than more metal-poor solar-type stars
\citep[e.g.,][]{san04,fis05}.  Likewise, metal-rich low-mass stars
also seem more likely to host giant planets \citep[e.g.,][]{joh09}.
Meanwhile, preliminary results indicate that near solar metallicity
there is not much of a correlation between solar-type host star
metallicity and the likelihood of hosting Neptune-mass planets
\citep[e.g.,][]{udr06,sou08,bou09}.  There are also trends in host stellar
mass, as higher mass stars are more likely to host giant planets than
lower mass stars \citep[e.g.,][]{joh07,joh10,bor11b}.

In \citet{sch10a}, we tentatively noted a hint that M dwarfs that host
low-mass planets are more likely to be metal rich than expected based
on random sampling from the field M dwarf population.  We explained
this correlation in the context of the core-accretion model of planet
formation \citep[e.g.,][]{pol96,ida04,lau04,ida05,hub05,mor09a,mor09b}.
In particular, we argued that a more metal-rich protoplanetary disk will
almost certainly have a higher surface density of solids.  That increased
solid surface density may enable the rapid formation of the several
Earth-mass embryo necessary to form the core of an ice giant and accrete
gas from the protoplanetary disk before the gaseous disk is dissipated.
That correlation might also indicate a lower limit on the amount of
solid material necessary to form planets.

In this paper, we determine whether {\it Kepler} EC host stars are
more metal rich than stars with no detected planets.  To that end, we
use a metallicity-sensitive color-color plot to examine the optical
and infrared properties of the sample of {\it Kepler} EC host stars
relative to a control sample of dwarf stars observed by {\it Kepler}
but that were not observed to host transiting planets.  We describe our
analysis procedures in Section 2, we discuss the results and implications
of our analysis in Section 3, and we summarize our findings in Section 4.

\section{Analysis}

We are interested in using the available photometric data for stars in
the {\it Kepler} field to determine if there is an offset in metallicity
between the {\it Kepler} EC host population and the population of stars
observed by {\it Kepler} but with no detected planets.  In particular,
we compare the mean $g-r$ color of {\it Kepler} EC hosts with the mean
$g-r$ color of stars observed by {\it Kepler} but with no detected planets
at constant $J-H$ color.  \citet{ive08} explored the potential of SDSS
photometry to produce accurate photometric metallicities for metal-poor
stars in the halo of the Milky Way.  Unfortunately, the accuracy of
those methods depends crucially on the availability of precise SDSS
$u$-band photometry.  Nevertheless, even when $u$-band data is lacking,
$g$,$r$, and $i$ photometry can still provide very useful photometric
metallicities \citep[e.g.,][]{an09b}.

Our problem is made easier because we are only interested in the relative
(not absolute) metallicities of the two populations.  At the same time,
unlike \citet{ive08} and \citet{an09b}, we have the added benefit of 2MASS
$JHK$ photometry in the {\it Kepler} field and apparent magnitude range.
As a result, we can use infrared colors (e.g., $J-H$) as a proxy for
effective temperature $T_{\mathrm{eff}}$.  The {\it Kepler} Input Catalog
\citep[KIC -][]{bat10b,bro11} provides reasonably accurate estimates
of $\log{g}$, so we can ensure that giant stars do not contaminate our
control sample of stars with no detected planets.  Stars with $M_{\ast}
\approx 0.7~M_{\odot}$ have main sequence lifetimes much longer than 10
Gyr, and because age has little effect on the colors of low-mass stars on
the main sequence, age differences between the two populations (if they
exist) are unlikely to affect our analysis.  Since we compare dwarf stars
with similar $T_{\mathrm{eff}}$ and $\log{g}$, the only stellar parameter
that can vary between the {\it Kepler} EC hosts and the control sample
is [Fe/H].  As metallicity is the only stellar parameter that can vary,
the best explanation for systematically red $g-r$ colors in the EC host
population relative to the control sample is that the EC host population
is enriched in metals.

\subsection{Color Offsets}

For each of the 997 stellar hosts of the 1,235 {\it Kepler} exoplanet
candidates announced in \citet{bor11b}, we obtain all available broadband
photometry \citep[including 2MASS $JHK$ from][]{skr06} and reddening
data from the KIC.  Simultaneously, we obtain equivalent photometry and
reddening data from the same sources for a random sample of 10,000 dwarf
stars from the $\sim\!\!150,000$ stars observed during Q1 and Q2 of the
{\it Kepler} mission with no candidate planets and KIC-based $\log{g}
> 4$.  This control sample is subject to the same selection effects
that were applied in the {\it Kepler} field to produce a list of stars
to search for transiting planets.  Consequently, differences in the
characteristics of the stars in the control sample and the stars that host
ECs are not related to the selection effects applied in the {\it Kepler}
field to identify a sample of stars suitable for transit observations.

Metal-rich stars typically have redder optical colors than solar
metallicity stars, as the forest of iron lines in the atmospheres of
metal-rich stars preferentially absorb blue photons.  At the same time,
there are other factors that affect the color of a star, most notably
reddening due to the interstellar medium, its age, and its evolutionary
state (e.g., dwarf or giant).

\begin{figure}
\plotone{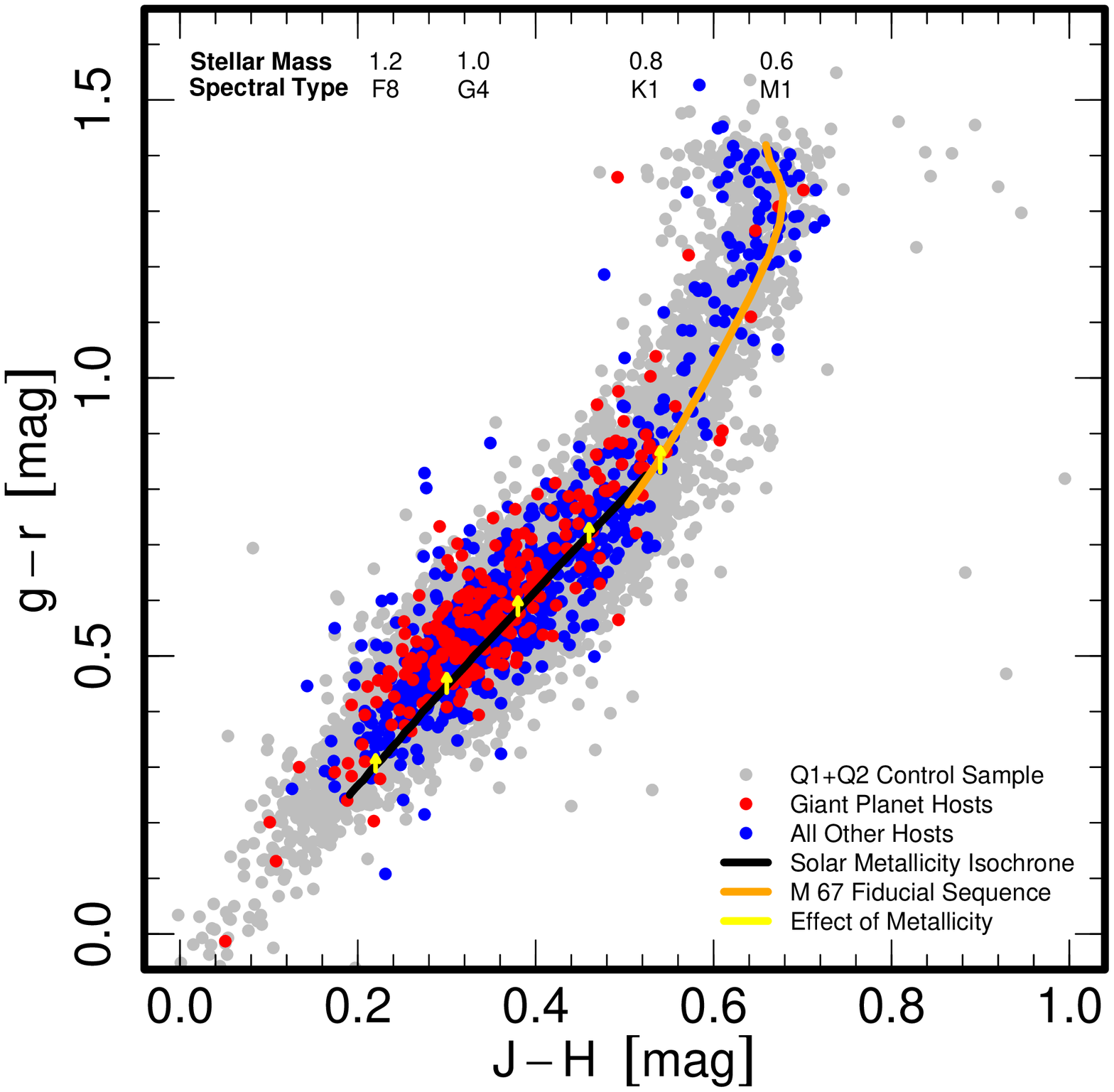}
\caption{{\it Kepler} exoplanet candidate (EC) host stars in a
$J-H$---$g-r$ color-color plot.  We plot in red stars that host at least
one giant EC (i.e., $R_p > 5~R_{\oplus}$), while we plot in blue stars
that host an EC system with no giant planets.  We plot in gray a control
sample of 10,000 stars randomly selected from the $\sim\!\!150,000$
stars observed in Q1 and Q2 of the {\it Kepler} mission that have no
detected ECs.  The black curve is a 2 Gyr, solar metallicity Padova
isochrone \citep{mar08,gir10}, and the orange curve is the M~67 fiducial
sequence from \citet{an08}.  We indicate with yellow arrows the affect
of increasing metallicity 0.2 dex in [Fe/H] on $g-r$ at constant $J-H$.
Note though that the yellow arrows do not connect stars of constant mass,
as a metal-enriched star will be about 5\% more massive than a solar
metallicity star at constant $J-H$ color.  We give approximate stellar
mass and spectral type as a function of $J-H$ color at the top of the
plot.  At $J-H \gtrsim 0.6$ (typical of late K dwarfs), giant ECs become
very rare relative to smaller ECs.  In other words, the {\it Kepler}
ECs confirm the correlation between host stellar mass and frequency of
giant planet occurrence \citep[e.g.,][]{joh07,joh10,bor11b}.  Moreover,
at $J-H \approx 0.62$ characteristic of late K dwarfs, the effect of
metallicity moves metal-rich stars to redder $g-r$ at constant $J-H$.
At that color, the population of K dwarfs that host ECs has a redder $g-r$
color than the control sample (see Section 2.3).  That offset is unlikely
to be explained by differential reddening between the small EC host
sample and the control sample, differences in age on the main sequence,
or the presence of giant stars in the control sample (see Section 2.2).
As a result, the sample of low-mass stars that host small ECs is likely
more metal rich than the control sample of stars in the same range of
$J-H$.\label{fig01}}
\end{figure}

In Figure~\ref{fig01}, we plot a $J-H$---$g-r$ color-color plot for the
control sample and the EC host sample.  We subdivide the EC host sample
into two sub-samples: those stars that host at least one giant EC (e.g.,
$R_p > 5~R_{\oplus}$) and those stars that host no giant ECs.  We use
Padova isochrones \citep{mar08,gir10} for a 2 Gyr stellar population
at [Fe/H] = -0.1 and 0.1 to determine the effect of metallicity on the
stellar locus in a $J-H$---$g-r$ color-color plot.  The yellow arrows show
the effect of increasing [Fe/H] by 0.2 dex on $g-r$ at constant $J-H$.
Note though that the yellow arrows do not connect stars of constant mass,
as a 0.2 dex [Fe/H] enriched star will be about 5\% more massive than a
solar metallicity star at constant $J-H$ color.  We do not use the Padova
isochrone redward of $J-H = 0.52$, as the theoretical stellar models and
atmospheres become unreliable after that point \citep[e.g.,][]{an09a}.
Consequently, we use the M~67 fiducial sequence from \citet{an08}
to illustrate the morphology of a solar metallicity population at red
$J-H$ color.  Age can also affect the color of main sequence star,
though near $J-H = 0.62$ age has little effect on the main sequence.
Giant stars have slightly bluer $g-r$ colors than dwarf stars at
constant $J-H$, so the presence of giant stars in the control sample
could in principle produce a color offset between the {\it Kepler} EC
host population and the control sample.  Fortunately, KIC-based $\log{g}$
estimates are accurate enough to ensure that giant stars are at most a
few percent of our control sample \citep{bas11,bro11}.

Two trends are immediately apparent in Figure~\ref{fig01}.  First, at
$J-H \gtrsim 0.6$ typical of late K dwarfs, giant ECs become very rare
relative to smaller ECs.  We interpret this as evidence in support of
the correlation between host stellar mass and likelihood of hosting a
giant planet \citep[e.g.,][]{joh07,joh10,bor11b}.  Second, at $J-H \approx
0.62$ dominated by late K dwarfs, ECs are preferentially found around the
reddest part of the stellar locus as defined by the control sample.
We quantify the significance of this latter trend in Figure~\ref{fig02}.

To determine the significance of the observation that low-mass stars
that host ECs are on average redder in $g-r$ than the stellar locus as
defined by the control sample, we bin both the control sample and each
sub-sample of EC hosts into 0.16 mag bins in $J-H$ color and compute
the mean and median $g-r$ color in that bin.  We estimate the error
in each mean and median by bootstrap resampling.  We find that at
$J-H \approx 0.3$ characteristic of solar-type stars, giant EC hosts
have preferentially red $g-r$ colors at the 4-$\sigma$ level relative
to the control sample.  On the other hand, at that same $J-H$ color,
the hosts of small ECs are indistinguishable from the control sample.
In addition, at $J-H \approx 0.62$ the stars that host small ECs are
4-$\sigma$ redder in $g-r$ than the control sample.  We summarize the
results of our calculations in Table~\ref{tbl-1}.

\subsection{Possible Reasons for Color Offsets}

We argue that the significant color offsets apparent in Figure~\ref{fig02}
and Table~\ref{tbl-1} are the result of the enriched metallicity of the
EC host sample.  Indeed, it reassuring that we identify the hosts of
giant ECs as significantly red in $g-r$ at constant $J-H$, as it is well
established that the stellar hosts of giant planets are preferentially
metal rich \citep[e.g.,][]{san04,fis05}.  Other possible explanations
for the observed offset are selection effects, differential reddening,
systematic age differences between EC hosts and the control sample,
and the presence of giant stars in the control sample but not the EC
host sample.

\subsubsection{Selection Effects}

The depth of a transit is proportional to $(R_p/R_{\ast})^2$, so it is
easier to identify the signal of transiting planet around a small star.
On the main sequence, both age and metallicity can affect the radius
of a star.  According to the Padova isochrones for a 2 Gyr population
\citep{mar08,gir10}, at $J-H = 0.3$ a star with [Fe/H] = 0.1 is about
5\% larger than a star with [Fe/H] $= -0.1$ and the same $J-H$ color.
At $J-H = 0.62$, a star with [Fe/H] = 0.1 is about 7\% larger than
a star with [Fe/H] $= -0.1$ and the same $J-H$ color.  According to
the Padova isochrones, at $J-H = 0.3$ a 5 Gyr old star is about 6\%
larger than a 2 Gyr old star at the same $J-H$ color.  At $J-H = 0.62$,
a 10 Gyr old star is about 2\% larger than a 2 Gyr old star at the
same $J-H$ color.  Consequently, if there were no correlation between
host metallicity and probability of hosting a planet or possible tidal
destruction of aged exoplanet systems, transit surveys would be more
likely to identify transiting planets around young, metal-poor stars.
As we show in Section 2.2.3, age on the main sequence has very little
effect on the $g-r$ color of a star at constant $J-H$, so these effects
are not likely to produce the color offsets we observe.

The $g-r$ color offset at constant $J-H$ color that we observe between
the {\it Kepler} EC host sample and the control sample is only meaningful
if the two samples were subject to the same selection effects.  As we
argued in Section 2.1, both samples were selected according to the same
{\it Kepler} target selection algorithm.  To verify that the two samples
are similar, we plot in Figure~\ref{fig03} six color-color plots and
two color-magnitude diagrams \citep[similar to][]{cov07}.  In all cases,
the control sample outlines the distribution of both the sample of giant
EC hosts and the sample of small-radius EC hosts.  The EC hosts that are
outliers in each panel may be unequal mass binary star systems or systems
with poorly-estimated reddening.  The {\it Kepler} IDs (KOI Numbers)
of a few of our most extreme color outliers are: 1161345 (984), 2446113
(379), 5356593 (644), 8162789 (521), 10470206 (335), 10514430 (263),
and 11465813 (771).

In short, the age or metallicity of the host star of an exoplanet does
not significantly affect the probability of detecting a transiting planet,
and our control sample is a reasonable control sample for our measurement
of relative color offsets.  We therefore conclude that selection effects
are unlikely to produce the color offsets we observe in Figure~\ref{fig02}
and Table~\ref{tbl-1}.

\subsubsection{Differential Reddening}

Reddening due to the interstellar medium can affect the observed color of
a star.  In our differential analysis, this will only affect our result
if the reddening of the control sample is systematically different
than the reddening of the {\it Kepler} EC host sample.  Fortunately,
Figure~\ref{fig04} indicates that control sample is subject to the same
reddening distribution as the EC host sample.  Like the EC host sample,
the control sample is also spread more or less uniformly over the {\it
Kepler} field, so any angular dependence on reddening is unlikely to
produce a significant color offset between the control sample and
the EC host sample.  We therefore conclude that differential reddening
between the control sample and the {\it Kepler} EC host sample is
unlikely to produce the color offsets we observe in Figure~\ref{fig02}
and Table~\ref{tbl-1}.

\subsubsection{Systematic Age Differences}

As stars age on the main sequence, their colors can evolve.  According to
the Padova isochrones, at $J-H = 0.3$ a 5 Gyr old star is about 0.1\%
redder in $g-r$ than a 2 Gyr old star at the same $J-H$ color.  At $J-H =
0.62$, a 10 Gyr old star is about 0.5\% bluer in $g-r$ than a 2 Gyr old
star at the same $J-H$ color.  Clearly, colors of stars do not evolve
significantly on the main sequence.

Tidal evolution in exoplanet systems can destroy close-in planets
\citep[e.g.,][]{gu03,mar04}.  Transit surveys are strongly biased towards
close-in systems, so it is possible that transit surveys may be biased
towards young stars, as tidal evolution may have already destroyed
planets that once orbited older stars.  Assuming that planets arrive in
the close proximity of their host star on nearly circular orbits and that
the timescale for eccentricity damping is very short, then the timescale
for tidal disruption is approximately \citep[e.g.,][]{ibg09,sch10b}

\begin{eqnarray}\label{eq1}
\tau_{\mathrm{dis}} & = & \frac{4}{117} \frac{a_{0}^{13/2}}{G^{1/2}}
\frac{M_{\ast}^{1/2}}{M_{p}} \frac{Q_{\ast}'}{R_{\ast}^5}
\left[1-(R_{\ast}/a_0)^{13/2}\right] \mathrm{,}
\end{eqnarray}

\noindent
where $a_0$ is the initial semimajor axis of the planet before tidal
evolution, $Q_{\ast}'$ is the specific dissipation function of the host
star, $G$ is Newton's gravitational constant, $M_p$ is the mass of
the planet, and $R_{\ast}$ is the radius of the host star.  If tidal
evolution efficiently destroys planets in the {\it Kepler} sample,
then the hosts of {\it Kepler} ECs may be preferentially younger than
stars in the control sample.  The median period of giant ECs is 13
days, while the median period of small-radius ECs is 10 days.  As we
plot in Figure~\ref{fig05}, the timescale for tidal disruption of such
systems is in excess of 100 Gyr.  At the same time, as we showed above,
age has little effect on the $g-r$ color of a main sequence star at
constant $J-H$.  We therefore conclude that systematic age differences
between the control sample and the {\it Kepler} EC host sample or tidal
evolution in EC systems are unlikely to produce the color offsets we
observe in Figure~\ref{fig02} and Table~\ref{tbl-1}.

\subsubsection{Presence of Giant Stars}

Giant stars have stellar radii $R_{\ast} \sim 100~R_{\odot}$, and because
transit depth is proportional to $(R_p/R_{\ast})^2$, the transit of a
planet with radius $R_p$ in front of a giant star with $R_{\ast} \sim
100~R_{\odot}$ is about 10,000 times harder to detect than the transit
of the same planet in front of a dwarf star with radius $R_{\ast}
\sim 1~R_{\odot}$.  As a result, {\it Kepler} is unlikely to identify
transiting planets around giant stars, so the sample of {\it Kepler}
EC hosts is very likely free of giant stars.

While we only used stars with KIC-based $\log{g} > 4$ in our control
sample, it is possible that a few percent of the stars in our control
sample are giant stars \citep{bas11,bro11}.  Giant stars have slightly
bluer $g-r$ colors at constant $J-H$ color than dwarf stars, so the
presence of a significant number of giant stars in our control sample
could explain the 0.08 mag $g-r$ color offset we observe at $J-H = 0.62$
between the sample of {\it Kepler} EC hosts and our control sample.

To investigate the affect of significant giant star contamination
in our control sample, we obtain all available broadband photometry
\citep[including 2MASS $JHK$ from][]{skr06} and reddening data from
the KIC for a sample of stars observed by {\it Kepler} during Q1 and
Q2 with KIC-based $\log{g} < 3$.  The stars in this sample are likely
giant stars.  Recall that the 10,000 stars in our control sample have
KIC-based $\log{g} > 4$, so we can add stars from the sample of likely
giant stars to our original control sample of likely dwarf stars to
create giant star contaminated versions of our our control sample.
In particular, we create four versions of the giant star contaminated
control sample with differing levels of contamination: 30\%, 10\%, 5\%,
and 3\% contimination.  To determine the $g-r$ color offsets expected at
constant $J-H$ color between a sample of 997 dwarf stars (the number of
{\it Kepler} EC hosts), we create a sample of 997 dwarf stars by randomly
selecting 997 stars from the original, not-intentionally contaminated
control sample.  As before, we compute the mean $g-r$ color in bins
of constant $J-H$ color for each of the five samples described above.
We plot the result of this calculation in Figure~\ref{fig06}.

\begin{figure}
\plottwo{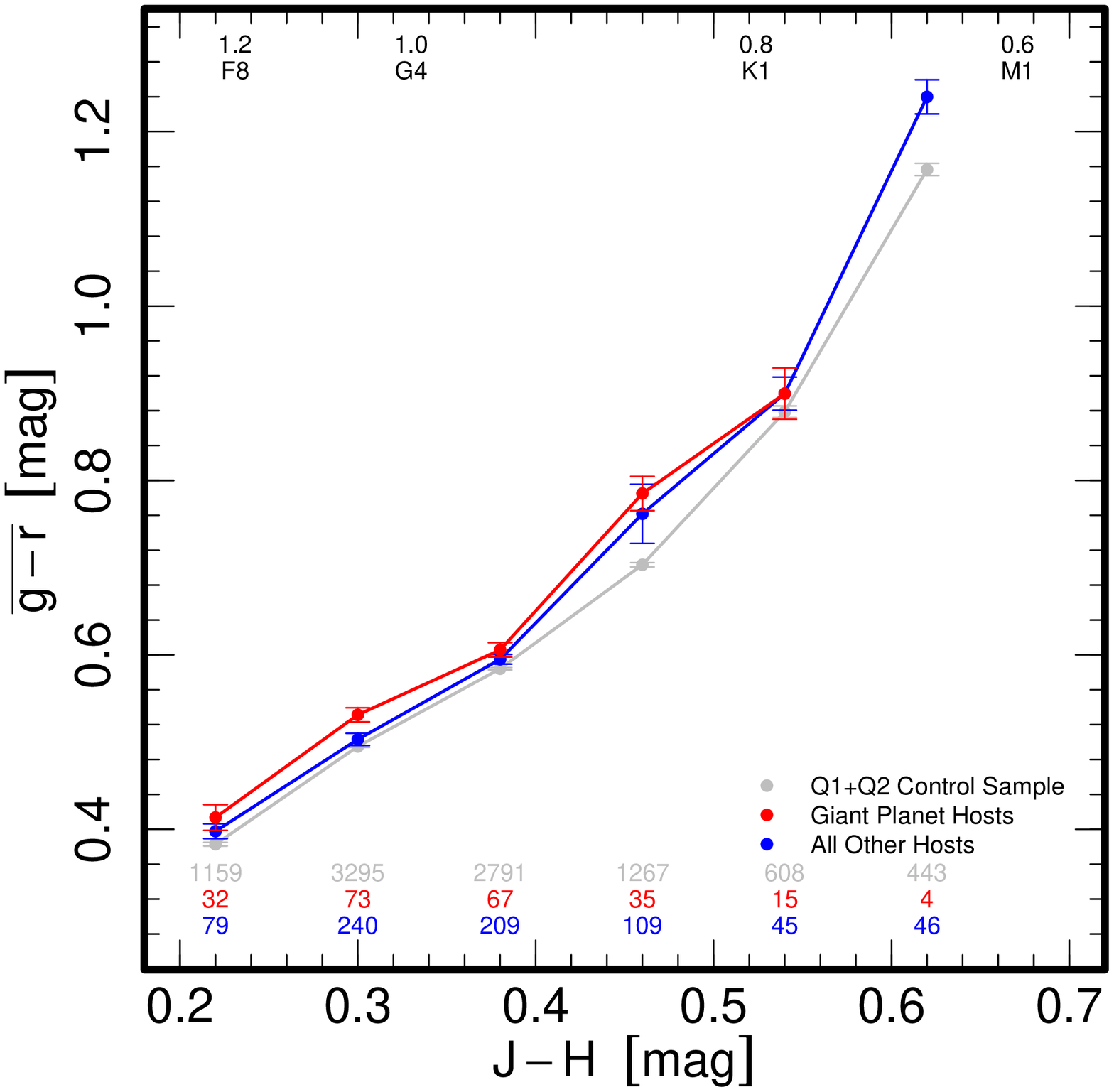}{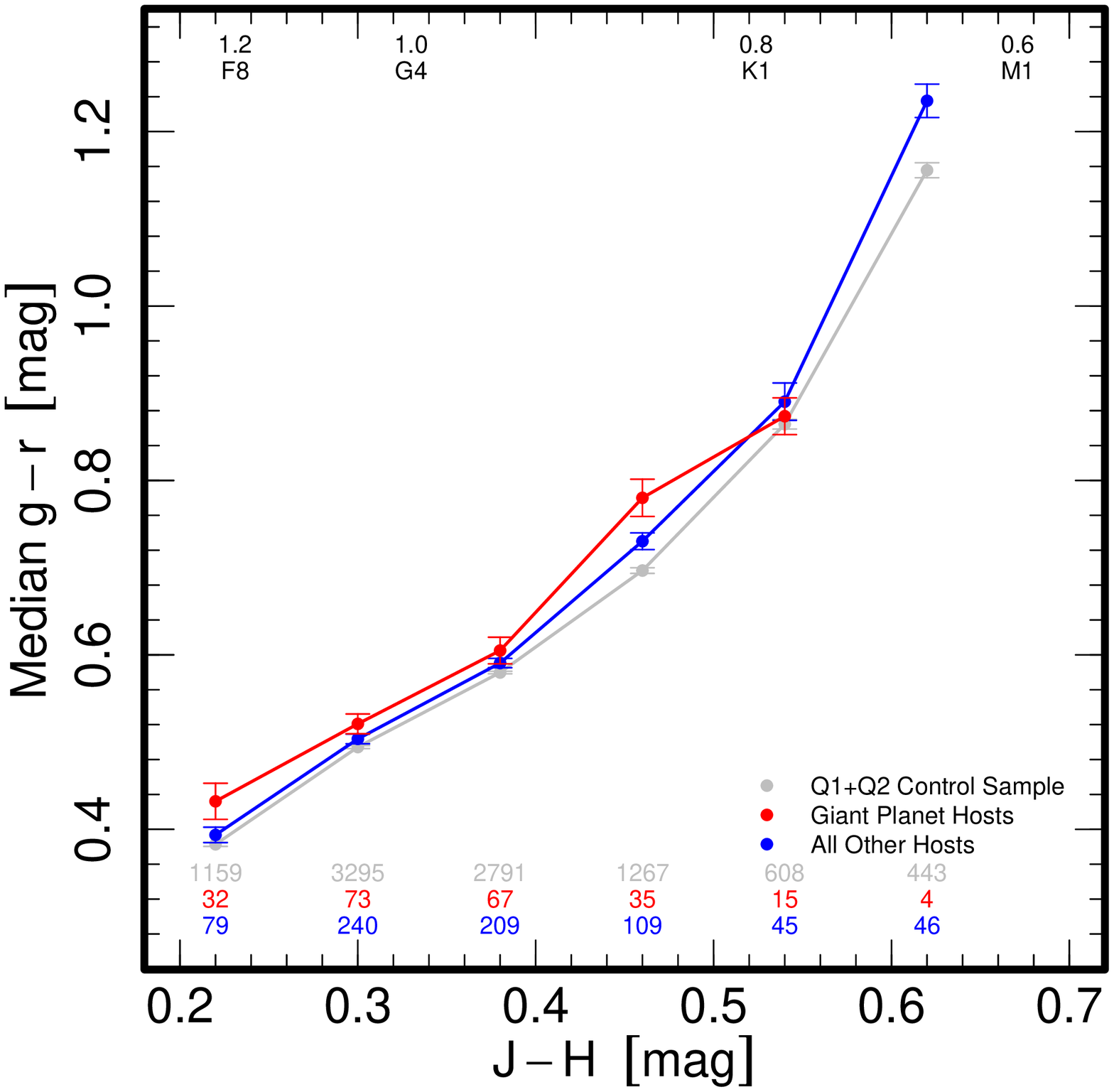}
\caption{Left: mean $g-r$ color as a function of $J-H$.  Right: median
$g-r$ color as a function of $J-H$.  We plot in red the result for the
hosts of giant ECs, in blue the result for hosts of systems without a
giant EC, and in gray the result for the control sample.  The error bar
on each point gives the 1-$\sigma$ error on our estimate of the mean
or median.   We give approximate stellar mass and spectral type as a
function of $J-H$ color at the top of the plot.  We give the number of
control stars, giant EC hosts, and small EC hosts in each 0.16 mag bin in
$J-H$ color at the bottom of the plot.  For solar-type stars, the hosts
of giant ECs have significantly redder $g-r$ at constant $J-H \approx
0.3$ (and therefore likely higher metallicity) than the control sample,
reproducing the observed correlation between giant planet occurrence
and host stellar metallicity \citep[e.g.,][]{san04,fis05}.  On the
other hand, there is no significant difference between the average
$g-r$ color of small EC hosts and the control sample, confirming
observations that for solar-type stars there is little correlation
between stellar metallicity and the likelihood of hosting a low-mass
planet \citep[e.g.,][]{udr06,sou08,bou09}.  At $J-H \approx 0.62$ however,
the hosts of small ECs are significantly redder (and therefore likely more
metal rich) than the stars in the control sample with no observed ECs.
We report the numerical values plotted above in Table~\ref{tbl-1}.
This result is a natural consequence of the core-accretion model of
planet formation and  confirms at much higher statistical significance
the tentative hint of that relation noted in \citet{sch10a}.\label{fig02}}
\end{figure}

\begin{figure}
\plotone{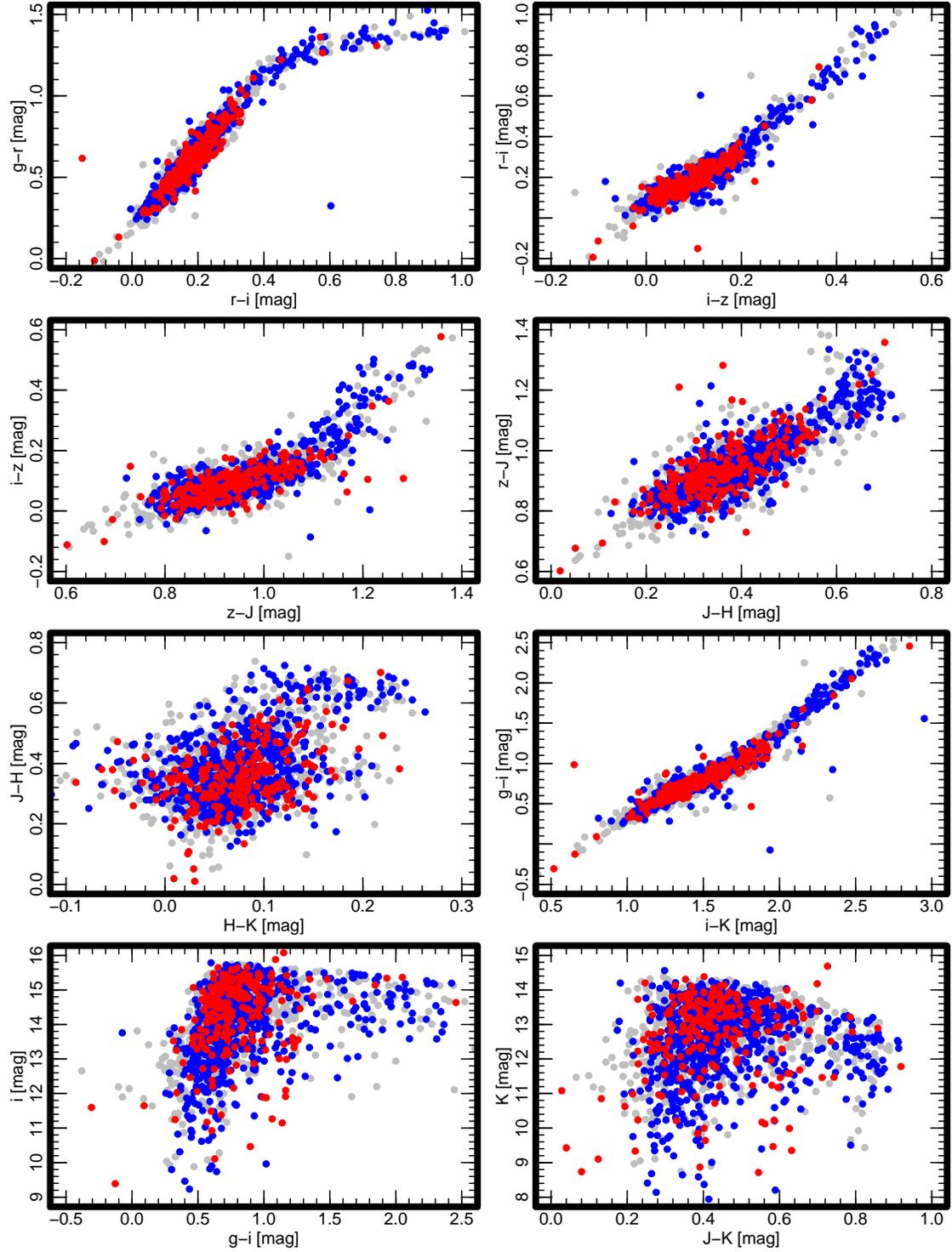}
\caption{Color-color and color-magnitude diagrams showing the distribution
of stars in the control sample in gray, stars that giant {\it Kepler} ECs
in red, and stars that host small-radius {\it Kepler} ECs in blue.  In all
cases, the control sample traces the same parameter space as the sample
of EC hosts.  The EC hosts that are outliers in each color-color plot
may be unequal mass binary star systems or systems with poorly-estimated
reddening.\label{fig03}}
\end{figure}

\begin{figure}
\plottwo{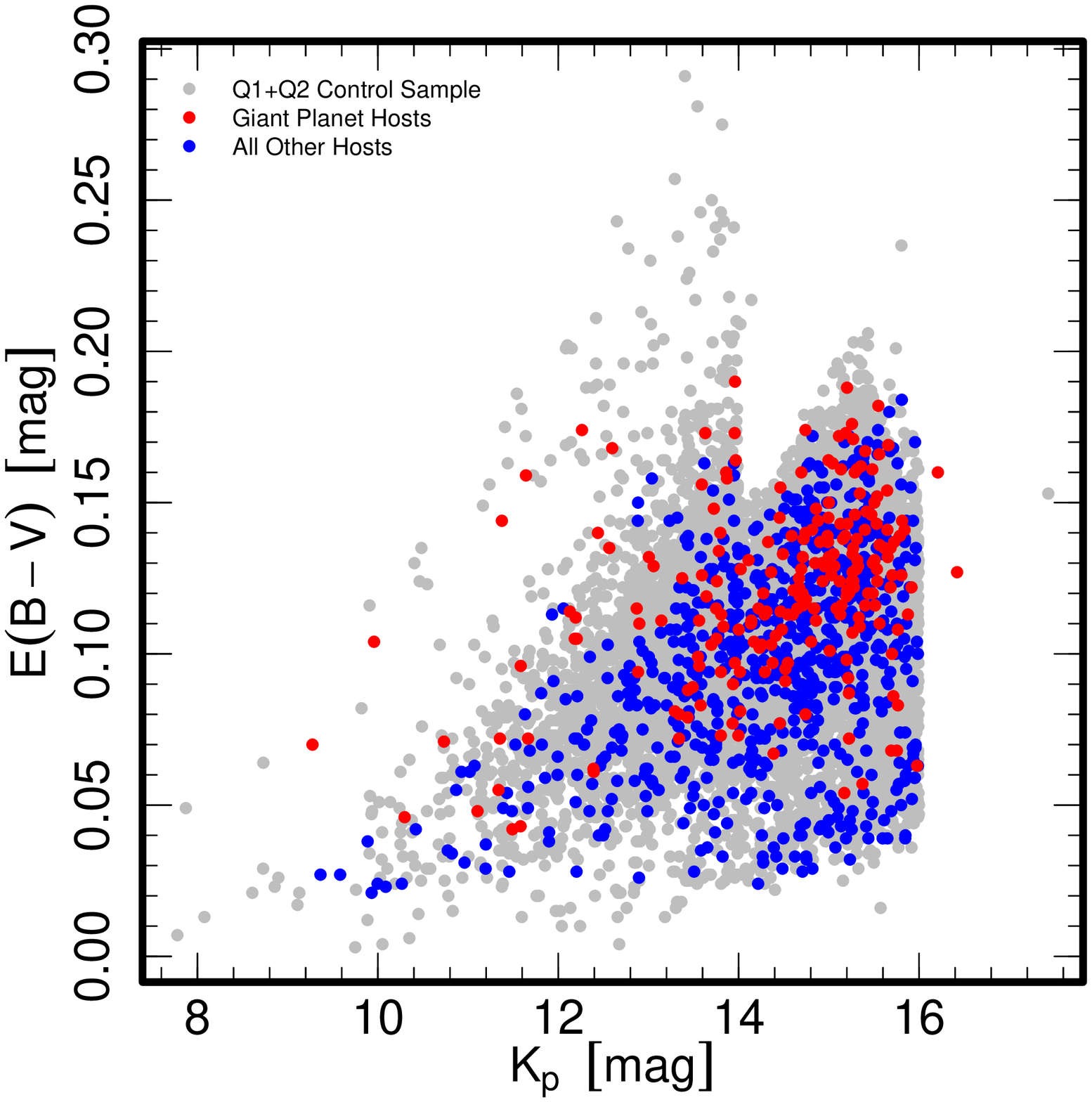}{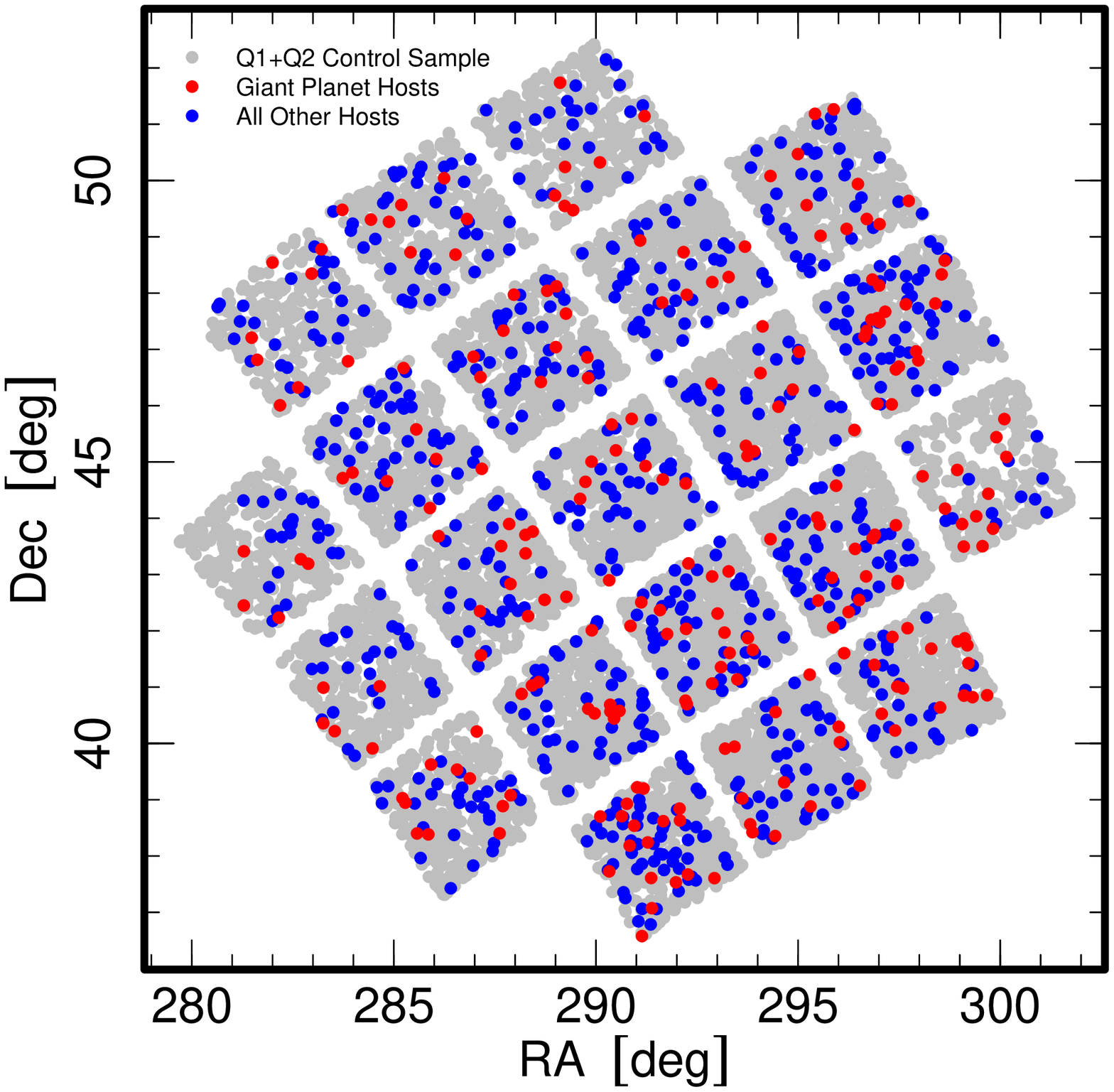}
\caption{Left: estimated color excess $E\left(B-V\right)$ of {\it Kepler}
EC host stars from the KIC as a function of apparent {\it Kepler}
magnitude $K_p$.  Right: equatorial coordinates of the stars used in
this analysis.  In both panels, we plot in red stars that host at least
one giant EC, while we plot in blue stars that host an EC system with no
giant planets.  We plot in gray the control sample.  The average color
excess of both the control sample and the giant EC host sample is 0.12,
while the sample of EC hosts with no giant planet has an average color
excess of 0.10.  The sharp edge in $E\left(B-V\right)$ at $K_p = 14$ is
likely a selection effect: apparently bright solar-type dwarf stars are
relatively rare, so they were included in the $\sim\!\!150,000$ stars
to be searched for transiting planets in Q1 and Q2 of the {\it Kepler}
mission regardless of reddening.  On the other hand, faint solar-type
dwarfs stars are numerous, so only the least reddened stars were included
in the sample to be searched for transiting planets.  The spatial
distribution of EC hosts on the sky is similar to the distribution of
stars in the control sample.  In short, the EC host sample is subject to
the same reddening as the control sample.  Therefore, any optical color
offsets between the control sample and the EC host sample are unlikely
to result from differential reddening.\label{fig04}}
\end{figure}

\begin{figure}
\plotone{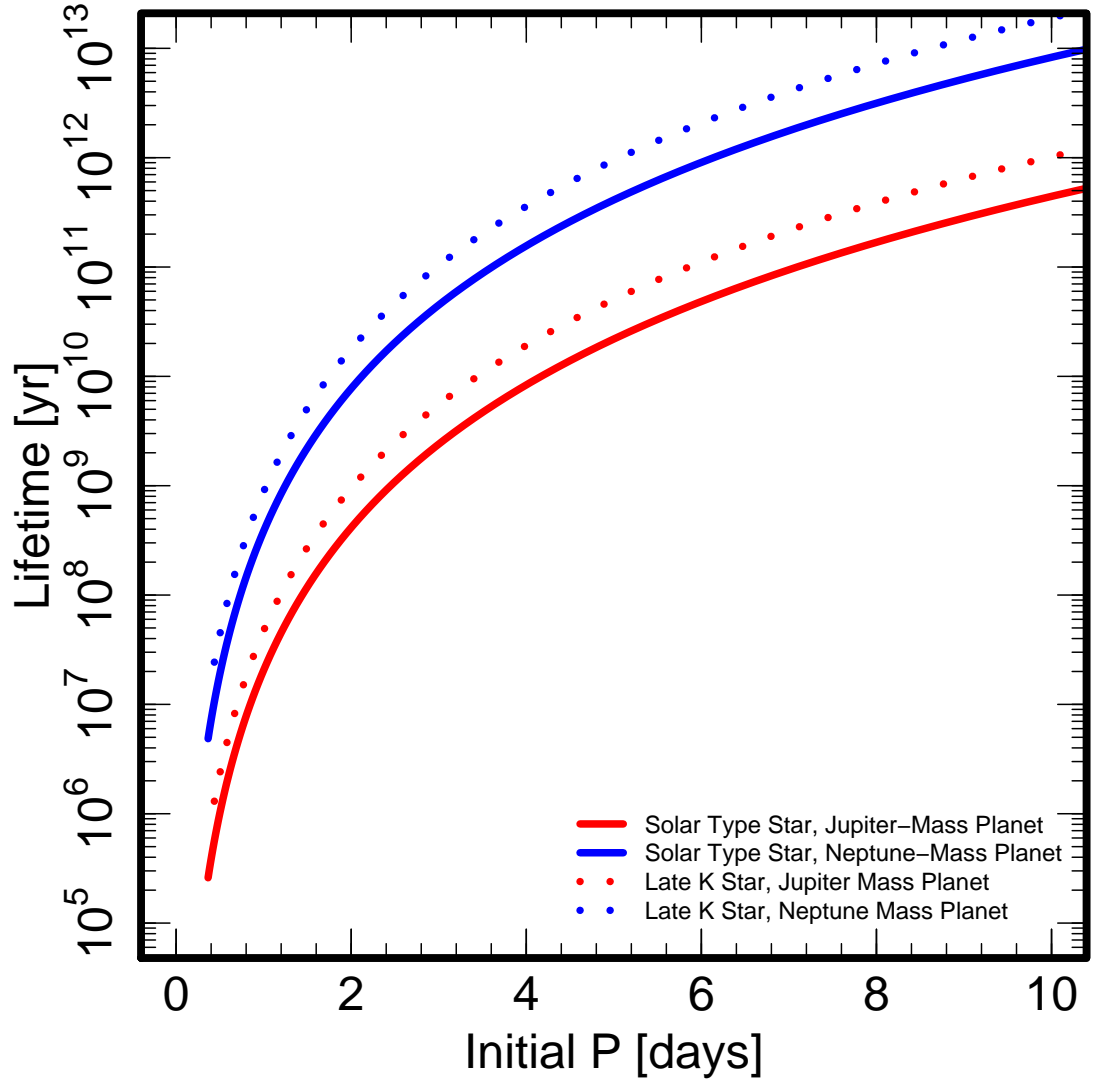}
\caption{Timescale for tidal migration to move a planet on an initially
circular orbit to one stellar radius, a reasonable approximation to the
timescale to tidal disruption.  In this case, we assume that $Q_{\ast}'=
10^6$, as \citet{sch10b} found that the average $Q_{\ast}'$ in the {\it
Kepler} EC host sample was $10^6 \lesssim Q_{\ast}' \lesssim 10^7$.
The median period of giant ECs is 13 days, while the median period of
small-radius ECs is 10 days.  As a result, tidal evolution does not
significantly affect the bulk of the {\it Kepler} EC, so there is no
reason to expect the hosts of {\it Kepler} ECs to be preferentially
young stars.\label{fig05}}
\end{figure}

\begin{figure}
\plotone{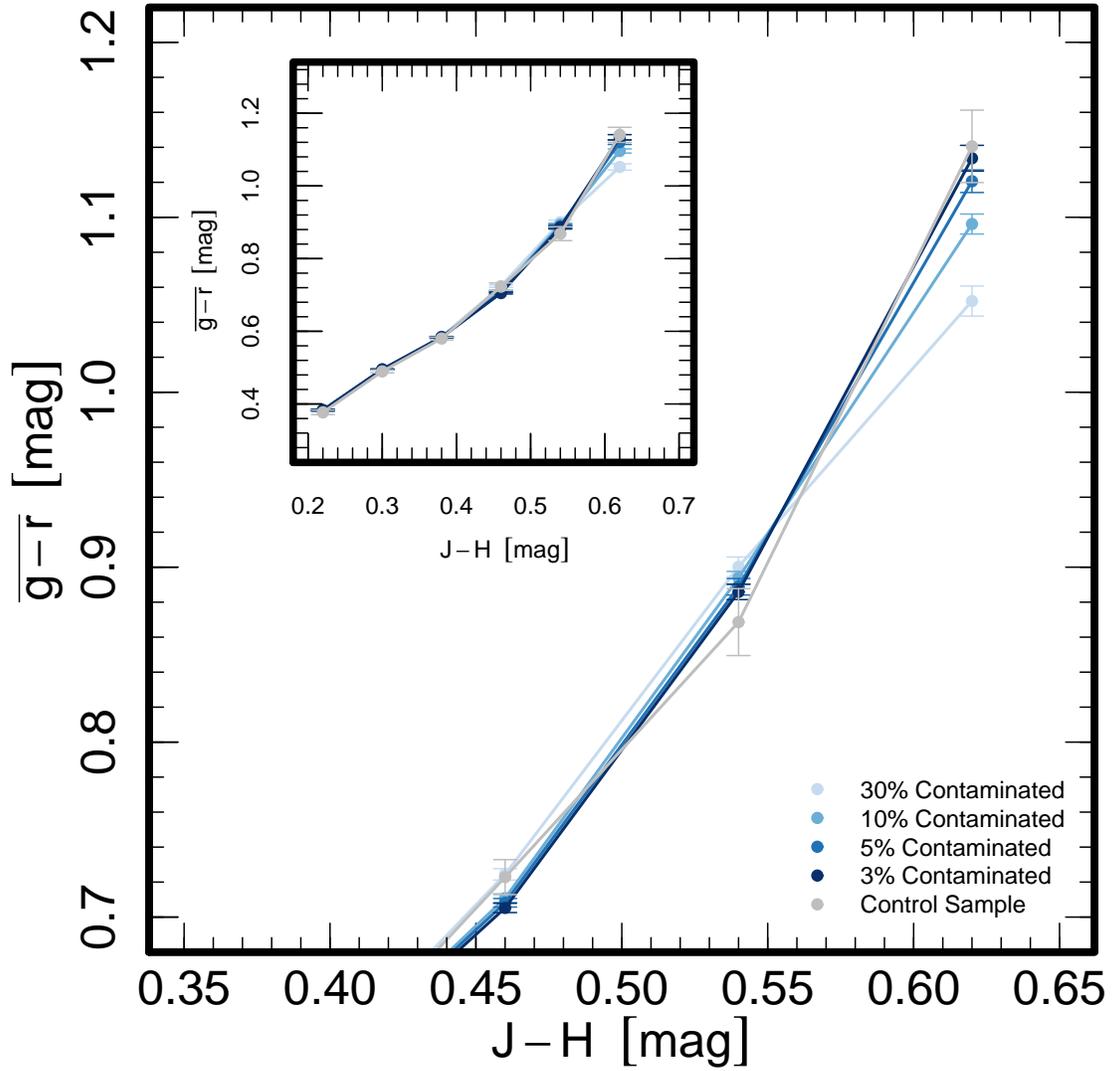}
\caption{Effect of giant star contamination in the control sample.
We find that between 10\% and 30\% of our control sample with KIC-based
$\log{g} > 4$ would have to be misclassified giants to explain the
the 0.08 mag $g-r$ color offset we observe at $J-H = 0.62$ between
the control sample and the hosts of small-radius {\it Kepler} ECs.
This level of contamination is at least a factor of a few larger than
the level of contamination expected in a sample of stars with KIC-based
$\log{g} > 4$ \citep{bas11,bro11}.  Contamination by giant stars cannot
explain the significant color offset we observe at $J-H = 0.30$ between
the control sample and the hosts of giant {\it Kepler} ECs.\label{fig06}}
\end{figure}

We find that between 10\% and 30\% of the stars in our control sample
of stars with KIC-based $\log{g} > 4$ would need to be giants to
reproduce the 0.08 mag offset in $g-r$ at $J-H = 0.62$ we observe
between the hosts of {\it Kepler} EC and our original not-intentionally
contaminated control sample.  The level of contamination is at least
a factor of a few larger than the level of contamination found in the
KIC by \citet{bas11} and \citet{bro11}.  Moreover, Figure~\ref{fig06}
indicates that a giant star contaminated control sample would have
redder $g-r$ color at $J-H = 0.54$ than control sample of dwarf stars
(very much like the likely hosts of {\it Kepler} ECs).  We do not observe
this effect in Figure~\ref{fig02}.  Indeed, the hosts of {\it Kepler}
ECs are redder than the control sample at $J-H = 0.54$.  At the same
time, the presence of giant stars has no effect on the $g-r$ color of
our sample at $J-H= 0.22, 0.30, \mathrm{or}~0.38$.  For that reason,
the presence of giant stars in our control sample cannot explain the
significant color offsets we observe between the hosts of giant {\it
Kepler} ECs and our original control sample.  We therefore conclude that
contamination of our control sample by giant stars is unlikely to produce
the color offsets we observe in Figure~\ref{fig02} and Table~\ref{tbl-1}.

\subsection{Metallicity Offsets from Color Offsets}

\begin{figure}
\plotone{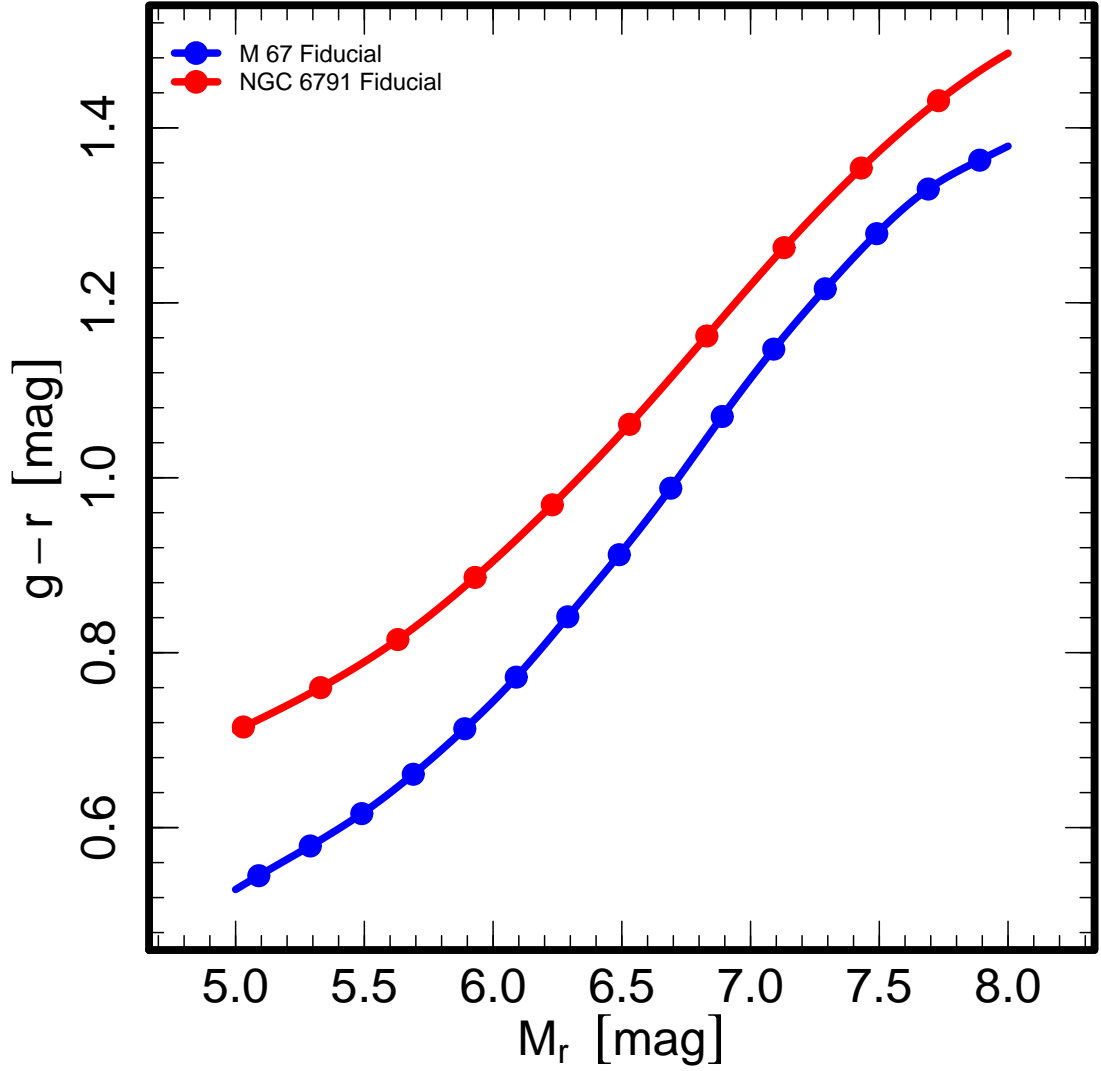}
\caption{Dereddened fiducial sequences for the open clusters M~67 ([Fe/H]
$= 0.0$ and age $\approx 4$ Gyr) and NGC~6791 ([Fe/H] = $0.4$ and age
$\approx 10$ Gyr) in the Sloan photometric system from \citet{an08}.
At $M_r \approx 6.8$ (corresponding to $J-H = 0.62$ characteristic of
late K dwarfs), the NGC~6791 fiducial sequence is 0.1 mag redder in $g-r$
than the M~67 fiducial sequence.  This offset is unlikely to be due to
the age difference between the two clusters, as the main sequence lifetime
of a late K dwarf is much longer than 10 Gyr.  In other words, at $J-H =
0.62$, every 0.1 dex increase in [Fe/H] corresponds to an increase in
$g-r$ of about 0.025 mag.\label{fig07}}
\end{figure}

We use two methods to transform the $g-r$ color offsets we observe in
Figure~\ref{fig02} and Table~\ref{tbl-1} into approximate metallicity
offsets.  First, we use dereddened fiducial sequences for the open
clusters M~67 ([Fe/H] $= 0.0$ and age $\approx 4$ Gyr) and NGC~6791
([Fe/H] = $0.4$ and age $\approx 10$ Gyr) in the Sloan photometric
system from \citet{an08} plotted in Figure~\ref{fig07}.  In this case,
the fiducial sequences show that at constant $6 \lesssim M_r \lesssim
7$, the metal-enriched NGC~6791 fiducial sequence is $\approx 0.1$
mag redder in $g-r$ than the solar metallicity M~67 fiducial sequence.
Though these two clusters have different ages, the stars with $6 \lesssim
M_r \lesssim 7$ have main-sequence lifetimes much longer than 10 Gyr
and therefore should still be on the main sequence in both clusters.
Consequently, the $\approx 0.1$ mag offset in $g-r$ between the two
fiducial sequences is likely a result of the $\approx 0.4$ dex offset in
[Fe/H] between the two clusters.  Accordingly, at constant $6 \lesssim
M_r \lesssim 7$ it seems that $\Delta\mathrm{[Fe/H]} = 0.1$ results in
$\Delta(g-r) \approx 0.025$.

To transform the $J-H$ colors of the points in Figure~\ref{fig02} and
Table~\ref{tbl-1} into $M_r$, we use a 2 Gyr solar metallicity Padova
isochrone.  To more precisely estimate the average $g-r$ offset between
the M~67 and NGC~6791 fiducial sequences, we calculate the difference
between the fiducial sequences at three points ($M_r$ = 5.9, 6.4, and 6.8
corresponding to $J-H$ = 0.46, 0.54, and 0.62) and average the result.
In this way, we find that in the interval $6 \lesssim M_r \lesssim 7$,
an offset of 0.1 dex in metallicity results in an offset of 0.035 mag
in $g-r$.  This method is only applicable to stars with $6 \lesssim M_r
\lesssim 7$, so in Table~\ref{tbl-1} we only give the metallicity offsets
indicated by this method for the three reddest $J-H$ bins in our analysis.

In order to transform a $g-r$ color offset into a metallicity offset for
more massive stars, we use Padova isochrones.  In particular, we use two
Gyr isochrones for two metallicities ([Fe/H] = -0.1 and 0.1) to determine
the change in $g-r$ color at constant $J-H$ attributable to a 0.2 dex
increase in metallicity.  We use the two isochrones to determine the
offset in $g-r$ at constant $J-H$ that can be attributed to metallicity,
then use that $g-r$ offset to convert the observed $g-r$ offsets in
Table~\ref{tbl-1} into metallicity offsets (extrapolating if necessary).

\begin{figure}
\plotone{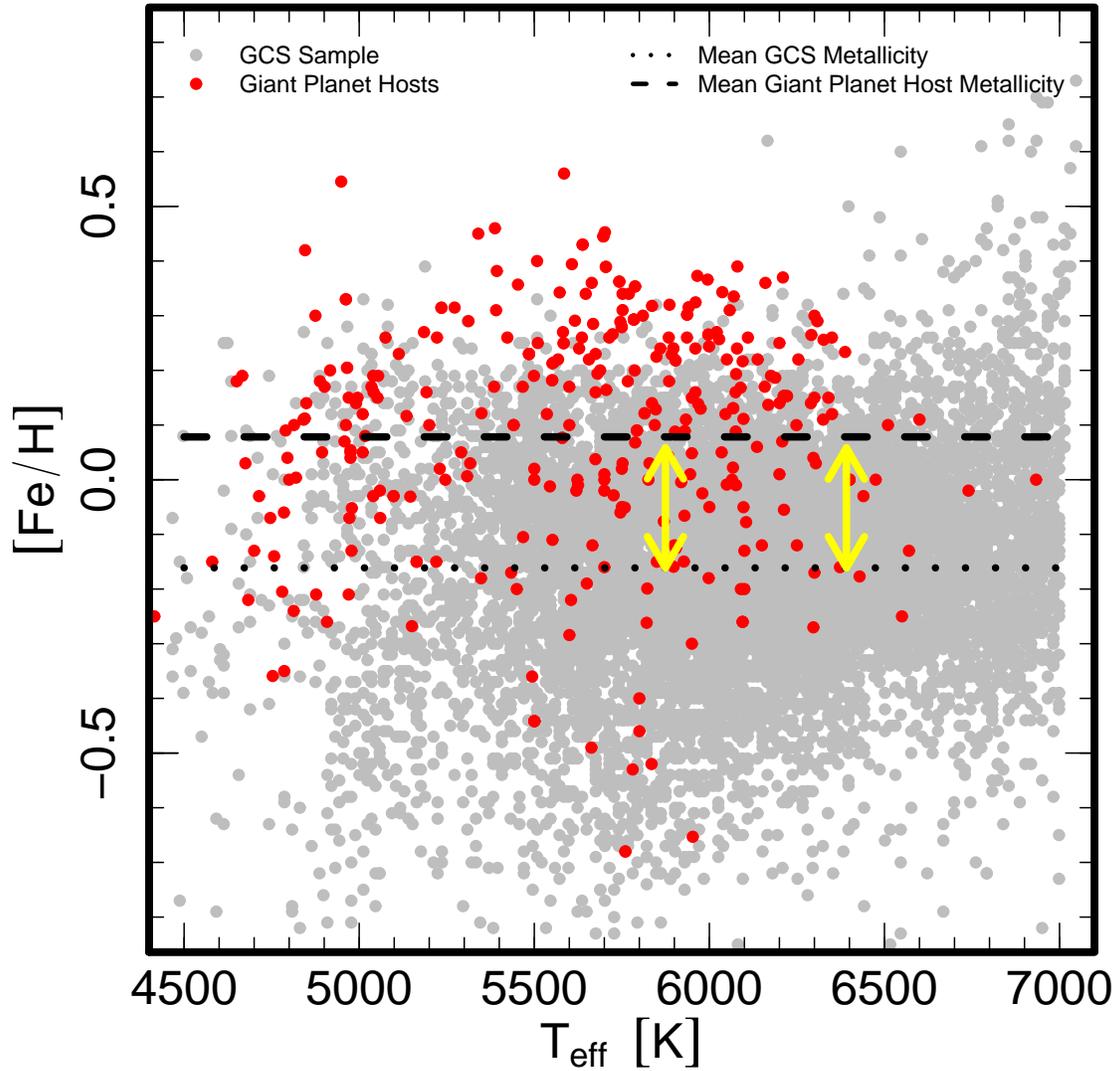}
\caption{Scatterplot of effective temperature $T_{\mathrm{eff}}$ and
metallicity for both a volume-limited ($d < 20$ pc) sample of stars from
the Geneva-Copenhagen Survey \cite[GCS -][]{holmb07,holmb09} and giant
planet hosts from \citet{wri11}.  We plot in gray the GCS stars and in
red the hosts of giant planets; the horizontal lines give the average
metallicities of the two populations.  The yellow vertical arrows
give the metallicity offsets suggested by the $g-r$ color offsets we
observe between the {\it Kepler} giant EC hosts and the control sample
at $J-H =$ 0.22 and 0.30 reported in Table~\ref{tbl-1}.  Indeed, though
our analysis is possibly imprecise and affected by several systematic
effects (e.g., differential reddening, age differences, or the possible
presence of giant stars in the control sample), we reproduce both
qualitatively and quantitatively the known metallicity offset between
solar neighborhood stars that host giant planets and those that do not
host giant planets.\label{fig08}}
\end{figure}

Interestingly, even in the presence of possible systematics in our
analysis (e.g., differential reddening, age differences, or the possible
presence of giant stars in the control sample), it is reassuring that
we recover the fact that solar-type stars that host giant planets are
metal enriched relative to the field population.  In Figure~\ref{fig08},
we plot $T_{\mathrm{eff}}$ and metallicity for both a volume-limited
sample of stars ($d <$ 20 pc) from the Geneva-Copenhagen Survey
\cite[GCS -][]{holmb07,holmb09} and confirmed giant planet hosts
\citep[e.g.,][]{wri11}.  The horizontal lines give the average
metallicities of the two populations.  The vertical arrows are at the
$T_\mathrm{eff}$ corresponding to $J-H$ = 0.22 and 0.30.  The arrows
are anchored at the mean metallicity of the GCS sample, and their length
corresponds to the metallicity offset we need to explain the differences
in $g-r$ color we observe in the giant planet bins at those $J-H$ colors
reported in Table~\ref{tbl-1}.  Reassuringly, the metallicity offsets
we compute based on $g-r$ color offsets are in quantitative agreement
with previous observations of metal-enhancement in giant planet hosts.
In other words, our analysis is quantitatively correct in the regime where
it is possible to compare with previous results.  That is suggestive that
our result for the low-mass stars that host small-radius {\it Kepler}
ECs is also reliable.

\section{Discussion}

We identified three properties of the {\it Kepler} ECs in Section 2:

\begin{enumerate}
\item
The population of solar-type stars that host giant ECs is 4-$\sigma$
redder by 0.04 mag in $g-r$ at constant $J-H = 0.3$ than a control sample
of stars observed by {\it Kepler} but with no detected planets.
\item
The population of solar-type stars that host small ECs is similar in
$g-r$ at constant $J-H = 0.3$ to a control sample of stars observed by
{\it Kepler} but with no detected planets.
\item
The population of low-mass stars that host small ECs is 4-$\sigma$
redder by 0.08 mag in $g-r$ at constant $J-H = 0.62$ than a control
sample of stars observed by {\it Kepler} but with no detected planets.
\end{enumerate}

The first observation is consistent with the known correlation between
solar-type host star metallicity and likelihood of hosting a giant
planet \citep[e.g.,][]{san04,fis05}.  The second observation supports the
tentative assertion that for solar-type stars in the solar neighborhood,
metallicity is only weakly correlated with the likelihood of hosting a
low-mass planet \citep[e.g.,][]{udr06,sou08,bou09}.  If the $g-r$ color
offset is the result of a 0.2 dex enhancement in [Fe/H] of the low-mass
hosts of small-radius {\it Kepler} ECs, then the third observation
supports the tentative assertion put forth in \citet{sch10a} that M
dwarfs that host low-mass planets are more likely to be metal rich than
expected based on random sampling of the field M dwarf population.

The observation that there is a correlation between host star metallicity
and the presence of low-mass planets around low-mass stars is expected
in the core-accretion model of planet formation.  In that scenario, the
cores of Neptune-mass planets (and possibly super-Earths) form and grow
to several Earth masses before their parent protoplanetary disk is fully
dissipated.  To first order, the mass of a protoplanetary disk scales
as $M_{\mathrm{disk}} \propto M_{\ast}$ and the fraction of the disk in
solid material scales as $f_{\mathrm{solid}} \propto Z_{\ast}$.  The total
amount of solids in a disk able to form the core of a Neptune-mass planet
is then $M_{\mathrm{solid}} \propto f_{\mathrm{solid}} M_{\mathrm{disk}}
\propto Z_{\ast} M_{\ast}$.  The protoplanetary disk in which the
solar system formed likely had a solid mass $M_{\mathrm{solid}} \sim
100~M_{\oplus}$ \citep[e.g.,][]{lis93}.  If the probability of forming
a planet is proportional to the total amount of solids in its parent
protoplanetary disk, a late K dwarf with $M_{\ast} = 0.7~M_{\odot}$ would
need to have a metallicity of [Fe/H] $=0.15$ to have the same chance
of forming a planetary system as a solar metallicity solar-type star.
As a result, it may only be the metal-rich late K dwarfs that have
protoplanetary disks with enough solid mass to form the several Earth-mass
core of an ice giant or super-Earth.  Though comparing stars at constant
$J-H$ color forces us to compare stars of slightly different masses,
the magnitude of this effect on the planet formation process is likely
an order of magnitude smaller than the effect of a 0.2 dex enhancement in
[Fe/H].

The correlation between the red optical color of late K dwarfs and
probability of hosting a small EC indicates that {\it Kepler} might
boost its yield of planets by shading their late K dwarf sample to
redder optical colors.  In particular, focusing on stars in the range
$0.5 \lesssim J-H \lesssim 0.7$ above the line

\begin{eqnarray}
g-r & = & 1.5 \left[(J-H)-0.5\right] + 1
\end{eqnarray}\label{eq01}

\noindent
might produce a larger sample of small ECs than a search that does not
use red $g-r$ color in its target selection.

\section{Conclusion}

We find that at $J-H=0.62$, low-mass stellar hosts of small-radius {\it
Kepler} ECs are 4-$\sigma$ redder in $g-r$ than a control sample of stars
with the same $J-H$ color but with no detected planets.  This result
is unlikely to be an artifact of reddening, age differences between the
two populations, or the presence of giant stars in the control sample.
Stellar models and fiducial sequences in the Sloan photometric system
of the Galactic open clusters M~67 and NGC~6791 indicate that a 0.2
dex enrichment in [Fe/H] of the EC host population can produce the
offset.  We suggest that small planets are preferentially found around
metal-rich low-mass stars \citep[confirming the tentative correlation
advocated in][]{sch10a}.  We also confirm that giant {\it Kepler} ECs
are preferentially found around metal-rich solar-type stars and that
the presence of small {\it Kepler} ECs orbiting solar-type stars does
not seem to be strongly correlated with host metallicity.

\acknowledgments We thank Tim Brown, Connie Rockosi, and Graeme Smith for
useful comments.  We also thank the anonymous referee for suggestions
that improved this manuscript.  This research has made use of NASA's
Astrophysics Data System Bibliographic Services, the Exoplanet Orbit
Database, and the Exoplanet Data Explorer at exoplanets.org.  Some of the
data presented in this paper were obtained from the Multimission Archive
at the Space Telescope Science Institute (MAST). STScI is operated by
the Association of Universities for Research in Astronomy, Inc., under
NASA contract NAS5-26555. Support for MAST for non-HST data is provided
by the NASA Office of Space Science via grant NNX09AF08G and by other
grants and contracts.  This material is based upon work supported under
a National Science Foundation Graduate Research Fellowship.

{\it Facility:} \facility{Kepler}

\clearpage

\clearpage

\clearpage

\clearpage

\clearpage

\clearpage

\clearpage

\clearpage

\clearpage
\begin{deluxetable}{cccccccccc}
\tablecaption{Optical Color Offsets and Equivalent Metallicity
Offsets\label{tbl-1}}
\tablewidth{0pt}
\tablehead{\colhead{$J-H$} & \colhead{$M_{\ast}$} & \colhead{$\Delta(g-r)$} &
\colhead{Significance} & \colhead{$\Delta$[Fe/H]}\tablenotemark{a} &
\colhead{$\Delta$[Fe/H]}\tablenotemark{b} & \colhead{$\Delta(g-r)$} &
\colhead{Significance} & \colhead{$\Delta$[Fe/H]}\tablenotemark{a} &
\colhead{$\Delta$[Fe/H]}\tablenotemark{b} \\
\colhead{(mag)} & \colhead{($M_{\odot}$)} & \colhead{(mag)} &
\colhead{($\sigma$)} & \colhead{(dex)} & \colhead{(dex)} & \colhead{(mag)} &
\colhead{($\sigma$)} & \colhead{(dex)} & \colhead{(dex)}}
\startdata
0.22 & 1.23 & 0.01 & 1.70 & $\cdots$ & 0.11 & 0.03 & 2.06 & $\cdots$ & 0.22 \\
0.30 & 1.05 & 0.01 & 1.14 & $\cdots$ & 0.05 & 0.04 & 4.44 & $\cdots$ & 0.22 \\
0.38 & 0.93 & 0.01 & 1.91 & $\cdots$ & 0.07 & 0.02 & 2.59 & $\cdots$ & 0.14 \\
0.46 & 0.85 & 0.06 & 1.72 & 0.17 & 0.39 & 0.08 & 4.11 & 0.23 & 0.54 \\
0.54 & 0.78 & 0.02 & 1.03 & 0.06 & 0.09 & 0.02 & 0.69 & 0.06 & 0.09 \\
0.62 & 0.73 & 0.08 & 4.00 & 0.24 & $\cdots$ & $\cdots$ & $\cdots$ & $\cdots$ & $\cdots$
\enddata
\tablecomments{Columns three through six correspond to the stellar hosts of
EC systems with no giant ECs, while columns seven through ten correspond
to the stellar hosts of EC systems with at least one giant EC.}
\tablenotetext{a}{Metallicity offset based on M~67 and NGC~6791 fiducial
sequences from \citet{an08}.  They are only applicable for stars with
$J-H \gtrsim 0.46$.}
\tablenotetext{b}{Metallicity offset based on Padova isochrones
\citep{mar08,gir10}.  They are only applicable for stars with $J-H
\lesssim 0.54$.}
\end{deluxetable}

\begin{thebibliography}{}
\bibitem[An et al.(2008)]{an08} An, D., et al.\ 2008, \apjs, 179, 326 
\bibitem[An et al.(2009a)]{an09a} An, D., et al.\ 2009a, \apj, 700, 523
\bibitem[An et al.(2009b)]{an09b} An, D., et al.\ 2009b, \apjl, 707, L64
\bibitem[Basri et al.(2011)]{bas11} Basri, G., et al.\ 2011, \aj, 141, 20
\bibitem[Batalha et al.(2010a)]{bat10a} Batalha, N.~M., et al.\ 2010a, \apjl,
713, L103
\bibitem[Batalha et al.(2010b)]{bat10b} Batalha, N.~M., et al.\ 2010b, \apjl,
713, L109
\bibitem[Batalha et al.(2011)]{bat11} Batalha, N.~M., et al.\ 2011, \apj, 729,
27
\bibitem[Borucki et al.(2010a)]{bor10a} Borucki, W.~J., et al.\ 
2010a, Science, 327, 977
\bibitem[Borucki et al.(2010b)]{bor10b} Borucki, W.~J., et al.\ 2010b, \apjl,
713, L126
\bibitem[Borucki et al.(2011a)]{bor11a} Borucki, W.~J., et al.\ 2011a, \apj,
728, 117
\bibitem[Borucki et al.(2011b)]{bor11b} Borucki, W.~J., et al.\ 2011b,
arXiv:1102.0541
\bibitem[Bouchy et al.(2009)]{bou09} Bouchy, F., et al.\ 2009, \aap, 496, 527
\bibitem[Brown et al.(2011)]{bro11} Brown, T.~M., Latham, D.~W.,
Everett, M.~E., \& Esquerdo, G.~A.\ 2011, arXiv:1102.0342
\bibitem[Bryson et al.(2010)]{bry10} Bryson, S.~T., et al.\ 2010, \apjl, 713,
L97
\bibitem[Caldwell et al.(2010)]{cal10} Caldwell, D.~A., et al.\ 2010, \apjl,
713, L92
\bibitem[Covey et al.(2007)]{cov07} Covey, K.~R., et al.\ 2007, \aj, 134, 2398
\bibitem[Dunham et al.(2010)]{dun10} Dunham, E.~W., et al.\ 2010, \apjl, 713,
L136
\bibitem[Fischer \& Valenti(2005)]{fis05} Fischer, D.~A., \& Valenti, J.\ 2005,
\apj, 622, 1102
\bibitem[Ford et al.(2011)]{for11} Ford, E.~B., et al.\ 2011, arXiv:1102.0544
\bibitem[Girardi et al.(2010)]{gir10} Girardi, L., et al.\ 2010, \apj, 724,
1030
\bibitem[Gu et al.(2003)]{gu03} Gu, P.-G., Lin, D.~N.~C., \& Bodenheimer,
P.~H.\ 2003, \apj, 588, 509
\bibitem[Haas et al.(2010)]{haa10} Haas, M.~R., et al.\ 2010, \apjl, 713, L115
\bibitem[Holman et al.(2010)]{holma10} Holman, M.~J., et al.\ 2010, Science,
330, 51
\bibitem[Holmberg et al.(2007)]{holmb07} Holmberg, J., Nordstr{\"o}m, B., \&
Andersen, J.\ 2007, \aap, 475, 519
\bibitem[Holmberg et al.(2009)]{holmb09} Holmberg, J., Nordstr{\"o}m, B., \&
Andersen, J.\ 2009, \aap, 501, 941
\bibitem[Howard et al.(2011)]{how11} Howard, A.~W., et al.\ 2011,
arXiv:1103.2541
\bibitem[Hubickyj et al.(2005)]{hub05} Hubickyj, O., Bodenheimer, P., \&
Lissauer, J.~J.\ 2005, Icarus, 179, 415
\bibitem[Ibgui \& Burrows(2009)]{ibg09} Ibgui, L., \& Burrows, A.\ 2009, \apj,
700, 1921
\bibitem[Ida \& Lin(2004)]{ida04} Ida, S., \& Lin, D.~N.~C.\ 2004, \apj, 616,
567
\bibitem[Ida \& Lin(2005)]{ida05} Ida, S., \& Lin, D.~N.~C.\ 2005, \apj, 626,
1045
\bibitem[Ivezi{\'c} et al.(2008)]{ive08} Ivezi{\'c}, {\v Z}., et al.\ 2008,
\apj, 684, 287
\bibitem[Jenkins et al.(2010a)]{jen10a} Jenkins, J.~M., et al.\ 2010a, \apjl,
713, L87
\bibitem[Jenkins et al.(2010b)]{jen10b} Jenkins, J.~M., et al.\ 2010b, \apjl,
713, L120
\bibitem[Jenkins et al.(2010c)]{jen10c} Jenkins, J.~M., et al.\ 2010c, \apj,
724, 1108
\bibitem[Johnson \& Apps(2009)]{joh09} Johnson, J.~A., \& Apps, K.\ 2009, \apj,
699, 933
\bibitem[Johnson et al.(2010)]{joh10} Johnson, J.~A., Aller, K.~M.,
Howard, A.~W., \& Crepp, J.~R.\ 2010, \pasp, 122, 905
\bibitem[Johnson et al.(2007)]{joh07} Johnson, J.~A., Butler, R.~P.,
Marcy, G.~W., Fischer, D.~A., Vogt, S.~S., Wright, J.~T., \& Peek, K.~M.~G.\
2007, \apj, 670, 833
\bibitem[Koch et al.(2010a)]{koc10a} Koch, D.~G., et al.\ 2010a, \apjl, 713,
L79
\bibitem[Koch et al.(2010b)]{koc10b} Koch, D.~G., et al.\ 2010b, \apjl, 713,
L131
\bibitem[Latham et al.(2010)]{lat10} Latham, D.~W., et al.\ 2010, \apjl, 713,
L140
\bibitem[Latham et al.(2011)]{lat11} Latham, D.~W., et al.\ 2011, \apjl, 732,
L24
\bibitem[Laughlin et al.(2004)]{lau04} Laughlin, G., Bodenheimer, P., \&
Adams, F.~C.\ 2004, \apjl, 612, L73
\bibitem[Lissauer(1993)]{lis93} Lissauer, J.~J.\ 1993, \araa, 31, 129
\bibitem[Lissauer et al.(2011a)]{lis11a} Lissauer, J.~J., et al.\ 2011a, \nat, 
470, 53
\bibitem[Lissauer et al.(2011b)]{lis11b} Lissauer, J.~J., et al.\ 2011b,
arXiv:1102.0543
\bibitem[Mardling \& Lin(2004)]{mar04} Mardling, R.~A., \& Lin, D.~N.~C.\ 2004,
\apj, 614, 955
\bibitem[Marigo et al.(2008)]{mar08} Marigo, P., Girardi, L., Bressan, A.,
Groenewegen, M.~A.~T., Silva, L., \& Granato, G.~L.\ 2008, \aap, 482, 883
\bibitem[Moorhead et al.(2011)]{moo11} Moorhead, A.~V., et al.\ 2011,
arXiv:1102.0547
\bibitem[Mordasini et al.(2009a)]{mor09a} Mordasini, C., Alibert, Y., \&
Benz, W.\ 2009, \aap, 501, 1139
\bibitem[Mordasini et al.(2009b)]{mor09b} Mordasini, C., Alibert, Y., Benz, W.,
\& Naef, D.\ 2009, \aap, 501, 1161
\bibitem[Pollack et al.(1996)]{pol96} Pollack, J.~B., Hubickyj, O.,
Bodenheimer, P., Lissauer, J.~J., Podolak, M., \& Greenzweig, Y.\ 1996,
Icarus, 124, 62
\bibitem[Santos et al.(2004)]{san04} Santos, N.~C., Israelian, G., \&
Mayor, M.\ 2004, \aap, 415, 1153
\bibitem[Schlaufman \& Laughlin(2010)]{sch10a} Schlaufman, K.~C., \&
Laughlin, G.\ 2010, \aap, 519, A105
\bibitem[Schlaufman et al.(2010)]{sch10b} Schlaufman, K.~C., Lin, D.~N.~C., \&
Ida, S.\ 2010, \apjl, 724, L53
\bibitem[Skrutskie et al.(2006)]{skr06} Skrutskie, M.~F., et al.\ 2006, \aj,
131, 1163
\bibitem[Sousa et al.(2008)]{sou08} Sousa, S.~G., et al.\ 2008, \aap, 487, 373
\bibitem[Udry et al.(2006)]{udr06} Udry, S., et al.\ 2006, \aap, 447, 361
\bibitem[Wright et al.(2011)]{wri11} Wright, J.~T., et al.\ 2011, \pasp, 123,
412
\end{thebibliography}
\end{document}